\DocumentMetadata{
     pdfversion=2.0,pdfstandard=ua-2,
      testphase={phase-III,firstaid,math,title}
    }

 \documentclass[sigconf,screen]{acmart-tagged}
\AtBeginDocument{%
  }

\setcopyright{cc}
\setcctype{by-nc-nd}
\acmDOI{10.1145/3757279.3785570}
\acmYear{2026}
\copyrightyear{2026}
\acmISBN{979-8-4007-2128-1/2026/03}
\acmConference[HRI '26]{Proceedings of the 21st ACM/IEEE International Conference on Human-Robot Interaction}{March 16--19, 2026}{Edinburgh, Scotland, UK}
\acmBooktitle{Proceedings of the 21st ACM/IEEE International Conference on Human-Robot Interaction (HRI '26), March 16--19, 2026, Edinburgh, Scotland, UK}
\received{2025-09-30}
\received[accepted]{2025-12-01}

\begin{document}

\title{Adding More Value Than Work: Practical Guidelines for Integrating Robots into Intercultural Competence Learning}


\author{Zhennan Yi}
\orcid{0000-0002-9039-2145}
\affiliation{%
  \institution{Indiana University Bloomington}
  \city{Bloomington}
  \state{IN}
  \country{USA}
}
\author{Sophia Sakakibara Capello}
\orcid{0009-0005-6186-0199}
\affiliation{%
  \institution{IUB \& UCF}
  \city{Bloomington}
  \state{IN}
  \country{USA}
}

\author{Randy Gomez}
\orcid{0000-0002-3191-6818}
\affiliation{%
  \institution{Honda Research Institute Japan}
  \city{Wako}
  \country{Japan}}

\author{Selma Šabanović}
\orcid{0000-0002-2553-854X}
\affiliation{%
  \institution{Indiana University Bloomington}
  \city{Bloomington}
  \state{IN}
  \country{United States}}

\renewcommand{\shortauthors}{Yi et al.}

\begin{abstract}
While social robots have demonstrated effectiveness in supporting students' intercultural competence development, it is unclear how they can effectively be adopted for integrated use in K-12 schools. We conducted two phases of design workshops with teachers, where they co-designed robot-mediated intercultural activities while considering student needs and school integration concerns. Using thematic analysis, we identify appropriate scenarios and roles for classroom robots, explore how robots could complement rather than replace teachers, and consider how to address ethical and compliance considerations. Our findings provide practical design guidelines for the HRI community to develop social robots that can effectively support intercultural education in K-12 schools.
\end{abstract}


\maketitle

\section{Introduction}

To prepare children for future participation in a globalized world, it is essential for them to develop intercultural competence (IC) \cite{unesco2015gced, byram2020teaching}. IC consists of intercultural knowledge, skills, and attitudes to effectively and appropriately interact with individuals from diverse cultural backgrounds \cite{byram2020teaching}. In educational practice, teachers often face challenges when promoting students' intercultural competence, due to linguistic and cultural barriers and inequitable access to resources \cite{chaika2024intercultural}. Social robots have potential to assist teachers as a complementary educational tool that can help address some of these barriers.



While the HRI community has drawn attention to designing social robots to support intercultural activities for students, most existing work has focused on multi-lingual inclusion (e.g., \cite{de2022engagement, kim2021designing}). Many existing studies 
have been conducted in controlled settings, such as labs or a separate room in school, which are isolated from real school settings or everyday routines (e.g., \cite{de2022engagement, kim2021designing, yi2025building, kim2024child}. More recently, HRI researchers have started working with teachers not only to design robots and related curriculum \cite{newbutt2022co, perez2024first}, but to create designs that provide opportunities for continued adaptation of robots to the evolving learning goals of teachers themselves \cite{gonen2025socially}. These works, while not directly addressing intercultural competence, highlight the importance of teachers' involvement in the successful deployment of educational technology.

In this study, we focus on the practical opportunities and considerations of using social robots to support children's intercultural competence. Specifically, we explore how robots can be meaningfully integrated into existing school contexts and routines, focusing on robots' added value within these settings, rather than creating stand-alone activities. 
Through two-phase workshops with a total of 17 K–12 teachers from different subjects and areas in the United States, we gained insights grounded in their actual classroom realities.
Using thematic analysis, we identify the added value of social robots to teachers, ways to embed them into existing relationships and routines in the classroom, and infrastructure needed to support robot use. 
We then propose a set of actionable guidelines to design social robots for intercultural activities in schools. Our work contributes to the HRI community by grounding the design of intercultural social robots in teachers' lived experiences, identifying how robots can be sustainable and educationally valuable, and laying the foundation for future in-situ implementations.

\section{Related Work}





\subsection{Social Robots in Intercultural Education}
Social robots can play a variety of roles in education \cite{belpaeme2018social}, supporting not only subject knowledge learning \cite{ligthart2023design, lee2022effects} but also the the development of students' different social emotional skills \cite{kewalramani2024scoping, marino2020outcomes, gordon2015can, park2017growing, almeida2023would}.
Intercultural competence is an important part of students' social emotional development. It consists of three components: intercultural knowledge, skills 
(skills of interpreting and relating, and skills of discovery and interaction) 
and attitudes 
(critical cultural awareness) 
\cite{byram2020teaching}. Frameworks such as \citet{byram2020teaching}'s Intercultural Communicative Competence Model and UNESCO's pedagogical guidance on global citizenship education \cite{unesco2015gced} highlight intercultural competence as central to preparing children for participation in a globalized world. Importantly, these competences are not automatic but need to be cultivated among children.
However, teachers face challenges when promoting them, including implicit bias and stereotyping, linguistic and cultural barriers, inequitable access to resources, resistance to change, and limited professional development and training \cite{chaika2024intercultural}. Social robots hold the potential to complement teachers in these areas. Compared to other technologies, robots offer additional personal and social dimensions \cite{belpaeme2018social}, which can be particularly useful for promoting the knowledge, skills, and attitudes needed for intercultural competence. 






Recent HRI research has explored the use of social robots to support intercultural activities. Most studies are situated within the context of ``translanguaging pedagogy'': using social robots as agents to support multilingual inclusion between students with different language backgrounds \cite{van2022social}. Examples include the use of social robots for storytelling-based local language learning among children with migrant backgrounds, which proved more effective than tablets for long-term learning \cite{de2022engagement}; and the use of a bilingual social robot to support collaborative activities between linguistically diverse child pairs, where bilingual component was found essential to engage them equally \cite{kim2021designing}.
More recent attempts use social robots as mediators in intercultural activities found that the presence of social robot mediators can help children from different countries open up and feel connected to each other \cite{yi2025building}; After playing robot-mediated conversational sessions, children pairs from different backgrounds had more liking, togetherness, and agreement behaviors than in the robot-mediated tablet sessions \cite{kim2024child}.


However, it is unclear how robot-involved intercultural activities can be implemented in educational contexts. Most of the existing work was either conducted in lab environments for a short-term interaction \cite{yi2025building} or in controlled conditions (e.g. in a separate small room in school \cite{de2022engagement}, in a closed conference room in the school \cite{kim2021designing}, during an after-school program with robots controlled by researchers \cite{kim2024child}), which are often disconnected and separated from daily school routines. While researchers could control the environment when in short-term interaction, 
real-life school classrooms have much more complicated and difficult conditions for robotic functioning \cite{woo2021use}. 
To design robot-supported intercultural activities that could be deployed sustainably in schools, it is necessary to understand current intercultural practices and the requirements for long-term adoption from practitioners' perspectives.

\subsection{Participatory Design with Teachers}

Teachers are key stakeholders in educational technology design as technology can greatly influence the educational practices for which they are responsible \cite{nordkvelle2005visions}. 
Yet teachers are still too often treated as implementers rather than co-creators of educational technologies \cite{tuhkala2021systematic, perez2017research}. While integrating robots into the classroom will inevitably influence their work, it remains unclear 
how teachers themselves envision the division of responsibilities between the robot and their own professional role. The review of field studies indicates that, when technologies are developed without considering teachers' roles, they are likely to not fit into the real classroom or become too difficult to sustain \cite{woo2021use}, highlighting the importance of engaging teachers as design partners rather than passive adopters.
Participatory design \cite{antonini2021overview, bodker2022participatory, chen2025robots} offers a human-centered method to approach these challenges by directly involving teachers in shaping how educational technologies could be designed, developed, and integrated into practice. 
Instead of positioning teachers as end-users who must adapt to systems, participatory design allows them to decide the role technologies play and the conditions for their use \cite{tuhkala2021systematic}, thus 
increasing their agency, and supporting faster and more effective adoption in practice \cite{gazulla2020co}.

\section{Methods}
We adopted the participatory co-design approach \cite{antonini2021overview, bodker2022participatory} by inviting K-12 teachers as co-designers.
We use the social robot Haru \cite{gomez2018haru}, a prototype that has been used in research on intercultural communication \cite{UNICEF-Haru}, as a design probe to help teachers brainstorm.
We presented a video of a \textit{Talking Room} session \cite{gomez2024design, yi2025building}, where Haru mediated communication between children from different countries, and a short video showing Haru leading a collaborative drumming activity. These examples provided teachers with an understanding of how social robots can be used in intercultural scenarios and demonstrated its interactive functions (conversation, gestures and movements, and facial animations). Notably, the focus was on creating school-based activities where social robots play a meaningful role, rather than designing the robot itself. 

We conducted \textbf{two phases of participatory workshops} with K-12 school teachers.
In the \textbf{first phase}, our aim was to have an overall understanding of teachers' practices in supporting children's general social-emotional skills and those related to intercultural competence, and explored the possibilities of robots-involved intercultural activities in school. 
We started by asking questions about teachers' current practices around soft skill development, intercultural goals and activities. After watching the two videos and discussing Haru's technical functions and potential roles, participants participated in two co-design activities: one focused on ideation, and another on identifying potential risks or challenges. Each session lasted 1.5 to 2 hours, and involved 1–3 teachers, for a total of 9 teachers from snowball sampling through  outreach to local schools and links. Two workshops were in person and the other were online using Zoom and Miro. In the \textbf{second-phase} workshop, we included some of the robot-involved activities designed by teachers from first-phase workshops to inspire discussions on the feasibility of certain aspects among additional teachers. After brief introduction of social robots and intercultural competence, participants were asked whether they have related intercultural teaching goals and their current practice. They watched the \textit{Talking Room} video, discussed how it could be adapted into their school, what might be inappropriate, and what other intercultural activities Haru could support. After that, they evaluated and discussed two designs from previous workshops. The workshop concluded with a brainstorming session and discussion of ``do''s and ``don't''s for HRI researchers and the robot development team. Each session lasted 1.5 to 2 hours. Each session of phase 2 workshops involved 2-3 teachers, for an additional total of 8 teachers recruited from Prolific. To mitigate potential fraud in the online study \cite{panicker2024understanding}, we implemented additional verification procedures by collecting participants' grade level, subject area, and years of teaching experience. These responses were checked for plausibility and reconfirmed during interviews. See Table \ref{demographic} for the demographic information of all participants.


Online sessions were automatically transcribed by Zoom, and in-person ones were transcribed manually, with all proofread by the first two authors.
We used reflexive thematic analysis to analyze our data\cite{braun2023doing}. 
We first familiarized ourselves with the data through repeated reading and note-taking to gain an in-depth understanding. Next, initial codes were generated inductively to capture meaningful features related to the research questions. These codes were then collated into potential themes, which were reviewed, refined, and defined to ensure coherence through several rounds of reflection and discussion among the authors.



\begin{table*}
    \caption{\textbf{Participants' Demographic Information}}
    \centering
    \begin{tabular}{p{0.02\linewidth} p{0.04\linewidth} p{0.02\linewidth} p{0.07\linewidth} p{0.05\linewidth} p{0.05\linewidth} p{0.44\linewidth} p{0.1\linewidth} p{0.03\linewidth}}
    \toprule
        ID &  Gender & Age & Experience & Grade & Ethnicity & 
        Subjects & School Types & State \\ \hline
        P1 & F& 45& 14 years& 7 \& 8 & Asian & Math, literacy, and science&Public	& IN\\ 
        P2 & F& 35& 10 years	& 4 \& 5& Asian &Data and science & Private& IN\\ 
        P3 & F& 56& 30 years& K–5& White 	&Art	&Public	&IL\\ 
        P4 & F& 28& 5 years&K–5&Asian  &Science& Public& NJ\\ 
        P5 & F& 31& 6 years& K& Asian  &Language	&Private&CA\\ 
        P6 & M& 28& 10 years&9–12& Asian 	& College/career; English as a second language	&Non-profit&CA\\ 
        P7 & M& 35& 1 year&K–8& Asian &	Chinese	&Charter &TX\\ 
        P8 & F& 28& 5 years&5& White 	&ELA/Social Studies (ESL)	&Public	&NJ\\ 
        P9 & F& 42& 20 years	&4 \& 8& White 	&Science	&Public	&IN\\ \hline
        P10 & F &55 & 26 years&	9–12& White &	Mathematics and Computer science	&Public	&NJ\\
        P11 & F& 31& 3 years	&3& White & 	English&	International &FL\\ 
        P12 & F& 29 & 8 years&	K–8& White &	ENL, Spanish	&Public&	IN\\ 
        P13 & F& 35& 10 years&	4& White 	&Math, Reading, Science, Social studies, Life skills	&Public	&IL\\ 
        P14 & F& 27& 5 years	&K–8& White & Science, Math, Literacy/Reading, Nutrition/Physical education&	Public	&CA\\ 
        P15 & F& 32& 8 years	&4& Hispanic 	&Reading/language arts	&Public	&TX\\ 
        P16 & F& 57& 11 years	&6–12& White 	&ELA, ESL, Reading intervention	&Public&CO\\ 
        P17 & F& 32& 7 years	&9–12& White &	Science&	Public Virtual	&WI\\ 
    \bottomrule
    \end{tabular}

    \label{demographic}
    \vspace{-1.5em}
\end{table*}

\section{Findings}

First, we present findings on the added value of having robots to support intercultural activities, focusing on the unique advantages and possibilities robots can bring in different scenarios. Second, we analyze how robots could be integrated into school settings effectively. Finally, we turn to broader infrastructure-level factors that influence the introduction and long-term use of robots in schools, with teachers' suggestions for addressing these challenges.

\subsection{The Added Value: Supporting Both Students and Teachers in Intercultural Activities}
\subsubsection{Supporting Multilingual Learning in Multiple Ways}

Our participants reported that accommodating students' varied language proficiency levels in class is challenging (P6, P7, P16, P17). 
P11, who taught in international school where students have varied knowledge of English, explained that there were always students who need extra individualized help, \textit{``because not everyone's English is going to be at the same exact level in the classroom.''} 

Teachers envisioned a social robot as a personal language guide for newcomer to learn and practice the local language. P10 pointed out the value of a robot bridging the communication gap in real time: \textit{
``[the student] was trying really hard to learn English, but I could see a case where I said something and he wasn't comprehending it, he could turn to the robot and say, `can you repeat that in French?' ''} P12, who taught English as a new language in a school with a large immigrant population where \textit{``a lot of newcomers do not speak a word of English''}, believed that the robot could speak their native language to make them \textit{``feel more comfortable and acclimate''}; gradually, the robot could pull back, and guide them to learn the targeted language.


During the language teaching process, teachers sometimes felt overwhelmed dealing with students' diverse needs in a limited class time. P7, who taught Chinese in a school where the majority of students were English speakers, expressed the difficulty of managing many students in one classroom: \textit{``my classroom has around 30 kids. So sometimes we cannot take care of each of them. It's impossible for that kind of custom setting.''}  For example, when practicing dialogue, teachers \textit{``have to repeat everything all the time, like asking the same questions, `what's your room like?' ''} (P7). 
Teachers envisioned social robots to share the workload by adopting flexible roles, including both teaching assistant demonstrating dialogue with teachers, and language practice buddy for students (P5, P6, P7), to reduce the repetitive work, \textit{``especially when sometimes [teachers] cannot go to every single group within the amount of time''} (P6).


Robots could be even more flexible in class. As P2 described, they \textit{``can be more flexible in terms of the different levels (it) can work with the students.''} They can act as role models and expert to guide students, but also as novices \textit{``just pretend that he doesn't know and ask students to explain, ask the students to become a teacher to explain the whole reasoning process.''} Depending on students' needs, teachers would like social robots to provide help or step-by-step guidance as reinforcement to individual students (P3, P9), or work with a small group of students (P16) as \textit{``an extra pair of eyes or ears''}, while teachers continued teaching the rest of the class.

\subsubsection{Mitigating Misunderstanding and Resolving Conflicts}

\label{conflict}

In intercultural interaction, students' difference in language and behavioral habits can easily lead to misunderstanding and even conflicts. P17 observed that sometimes innocent question like \textit{``why do you say that word that way?''} could actually be offensive to somebody else. P16 taught in a refugee program, which had a substantial increase over the past two years in students with very limited English proficiency and social skills. P16 shared that small comments, like \textit{``speak English!''}, could convey different meanings, then cause misunderstandings and lead to group fights in her school. As she put it, \textit{`` When the [anonymized student group] first arrived, there were some gang \footnote{The term ``gang'' may carry unconscious bias and is perceived as a racialized term \cite{whydont}.} fight things going on outside the classroom, where they arranged it without the teachers knowing. They met on the field, and just started attacking each other... When that boy said, `Hey, speak English', I don't think he meant anything by it; he was just using the words he knew. ''} 

Teachers found it tough explaining the misunderstanding to students partially due to language gaps, as commented by P16: \textit{``because I can't speak all the languages, and even my students can't read all of their languages either. ''} P16 even tried to learn students' language but faced great difficulties.
In teachers' perspective, robots \textit{``have more information, more resources''}, so they are likely to provide \textit{``more accurate''} language communication, with less time than a human teacher would need to achieve the same (P14). 

More importantly, teachers stated that dealing with conflicts in intercultural activities required extra sensitivity, since teachers themselves are human with cultural identity. Robots, compared to teachers, were seen as more neutral without personal or cultural bias when addressing misunderstandings. As P16 described, \textit{``if I try to do the conflict resolution... then there's my own personality also getting involved, but Haru wouldn't have that. They wouldn't take offense as much from this robot.''} They believed this neutral standpoint can make students more receptive during conflict resolution. Similarly, P15, who worked at a school with a predominantly Hispanic student population and who also identified as Hispanic, noted that, bringing in a third-party who is less familiar to the students can help reduce bias and increase their receptiveness: \textit{``Especially when it comes to mediating conversations or conflict resolution, I feel like sometimes the students don't want to hear from me... They spend so much of their day with me. It'd come better from a different origin. They could respond better to having it come from someone else.''}
In addition, while power dynamics inherently shape the teacher-student relationship, robots, thanks to their unique appearance and functions, can serve as a relatively lower-power presence in the classroom. As P2 said, \textit{``children would not take Haru as serious as they think about teachers... Haru is cute, children may take it as their friends and open up more.''} With a neutral standpoint, students are likely to feel more comfortable sharing their problems because they \textit{``are not necessarily worried that hardware is going to tell them that they're wrong or something''} (P8).

\subsubsection{Encouraging Inclusion and Equitable Participation}

How to ensure equitable participation of students from diverse cultural backgrounds is always a challenge to teachers in intercultural activities. P1, as a ESL speaker working in English-dominant classroom environments, noticed that students who were native English speakers in classroom usually participated more in group activities: \textit{``I can see from my classroom that has students from many countries. It might be because of the language limitation for students from other countries, I can see when they need to work in group, usually [anonymized cultural group] student did almost everything. And the student from other country, they had no chance to speak or to do something... just sit there and be quiet.''}
In another case, having too many students in the classroom who speak the same language, even if it is not the school's primary language, can cause an imbalance and may even leave the teacher excluded. As reported by P16: \textit{``I have so many [anonymized language (AL)] speakers in the class that they can start dominating the discussions in [AL], and I always thought my [AL] was good, because I grew up around it. But I didn't realize that I had accidentally only understood [a subtype of the AL]... so they'll just start talking and saying things, and I'll catch some words. But for the most part, I really can't follow what they're saying. ''}

Imbalanced participation is very likely to occur among students from diverse cultures who are likely to have diverse personalities and skill levels. It is also where teachers see the possibilities social robots could bring in.
During group activities, robots were imagined as group activity mediators that increase engagement and encourage participation. Especially in a group with students having different language skills and different levels of extraversion in social interactions, teachers highlighted that robots could \textit{accommodate different levels} (P8, P11) – encourage quieter students to join in by saying \textit{``hey, what are you thinking about?''} (P12), while mediating dominating voices to support turn-taking (P2, P8, P11). By providing feedback like \textit{``compliments''} (P1), having some personalities like \textit{``being a little bit funny, silly to make people laugh and be more engaged''} (P2), social robots have the potential to sustain engagement and motivate students. Robots were also envisioned as supportive listeners.
When students feel overwhelmed about talking to people, it provides them a temporary safe retreat to express without social pressure, as \textit{some shy kids may find it more helpful and easier to speak to a robot than an actual human} (P12, P7).

\subsubsection{Adopting Context-based Cultural Knowledge}

Teachers mentioned that students were open to and enthusiastic about learning from other cultures, particularly when they could find the connections to their own. However, teachers are not always equipped with such intercultural knowledge. 
For example, P8, who taught at a school where the majority of students had Spanish as their first language, shared a cultural gap in math method learning she noticed in her school: \textit{``
In a classroom here, we have a lot of students from [anonymized countries], and they have so many different ways of doing math problems that we don't teach here, or we weren't taught how to do them. Then when they're being assessed, they get told more to do the methods that are being taught here.''} 

A notable advantage of social robots, as a kind of technology, is their ability to access and provide vast resources that go beyond the capabilities of individual teachers.
In this case, teachers in one education system were only trained to teach certain methods, while social robots could bridge this gap with its powerful ``learning ability'', as P8 put it: \textit{`` I'm putting a lot of faith in Haru to understand the lesson versus the United States teachers... I think Haru will be able to understand it more easily than a teacher whose brain has been coded to teach a certain way.''} P13 highlighted, robots like Haru \textit{``have more information on a certain topic than teachers would, or teachers may have to do more research to understand''}. 
Leveraging this advantage, teachers envisioned using social robots as an addition to class to give more cultural backgrounds of their class topics, such as introducing cultural background of arts to encourage two-way conversations and reflections (P3).
Whenever students have any questions when learning other cultures, they could also directly ask the robot (P15).

Importantly, social robots like Haru could not only mediate discussion on intercultural knowledge, but their unique identity could be learning resources. As mentioned by P15, robots' unique backgrounds, which differ from real people's, can help students learn different cultures in an intuitive way: \textit{``
    Haru could focus on its background, like where it comes from. My students would be very in tune, and that would engage them, because they could be like, `Haru is different. Its background is not like ours, it is...' 
    Whereas they see me, they'll be like, `oh, Ms. [anonymized name], she's like us, she's [anonymized race].' But Haru isn't. ''} Thus robots could add engagement while fostering intercultural understanding.

\subsection{Embedding Robots into Classroom Dynamics}

In this section, we present our findings on how robots can be embedded into school settings, fitting within existing relationships and dynamics. Specifically, robots should (1) engage with students in culturally and developmentally appropriate ways, (2) be positioned with clearly defined boundaries and rules, and (3) integrate into teachers' existing practices and workflows.

\subsubsection{Developmental and Cultural Appropriateness}
When engaging with students in intercultural settings, robots must have their behaviors appropriate both culturally and developmentally.






First, make sure that the content is culturally or contextually appropriate. The content and language robots use should be suitable and respectful for the cultural groups and classroom environment. This includes no inappropriate topics or words to come up (P14), and no inappropriate languages such as \textit{comments or jokes that could be taken out of context} (P11), \textit{toilet jokes and any kind of inappropriate humor} (P3, P11), \textit{sarcasm} (P10). 
Second, robots should be able to adjust their interaction style to match students' developmental levels, learning preferences and needs in intercultural context. For different age groups, there are different preferences regarding the ways to engage in intercultural activities. 
Elementary kids have short attention spans (P15); Young children rely on many different cues to understand things, like image, visual signals, physical signal, non-verbal signals (P5), and they have many questions about everything (P4); And kids love stories and videos (P13). Especially \textit{``while human use different tones and emphases, robot say everything in a similar way,''} having those interactive components would be extremely helpful for students in a multi-cultural classroom who \textit{``are not native English speaker and are still learning it''} (P5) or when students struggle to understand (P12). For older kids in 4th or higher grades, they are not likely to take the activity serious, sitting down and having a social emotional learning session with robot as 1st and 2nd graders do, but simply letting them talk about feelings and experiences would be meaningful (P9, P14). Having different conversation styles and a different breadth to the questions would be helpful for students from different levels (P17). In addition, teachers raised the concerns for special education students (P13, P15). 
For nonverbal kids using assistive devices who struggle to sit and wait, they need some simple and instant activities like stories (P13).

\subsubsection{Clear Boundaries and Rules of Interactions}

When robots are introduced into the existing dynamics between teachers and students, they should be positioned with clearly defined boundaries and rules.
Teachers emphasized that social robots should not be merely a source of entertainment for students, instead, their use must be tied to explicit learning goals. Without such boundaries, students can easily get off-task, such as starting \textit{``getting into discussing something they saw on YouTube the other day''} (P11), or simply playing with robots by asking random funny questions (P7). As P11 put it, \textit{`` There's fun in learning, but if it's 99.9\% just fun, that puts learning outside of the lesson. Then there needs to be a bit more balance there. ''} This highlights the importance of defining the scope of robot interactions and balancing engagement with intercultural learning.
This can be achieved by setting expectations, and limit the usage of robots on robots' response scope, interaction time, and students' access.

Teachers suggested that before students engage in robot-involved intercultural activities, it is necessary to manage students' expectations of the interactions. For example, P3 noted the importance of giving students an introductory session to explain the goal, the usage of robots and expected student behaviors, to make sure that\textit{``it can fully function, and become a useful tool for us.''}



Another important requirement is to limit robots to response only to restricted scopes. As suggested by P15, instead of responding to every questions from students, robots should not answer those \textit{``have nothing to do with the purpose of what the robot is intended to do, which is to mediate intercultural activities and provide that social-emotional aspect of learning.''} To successfully use robots for students' intercultural development, robots should have the \textit{filtering} ability. According to P8, the robot should respond when a student asks for clarification about intercultural content but ignore unrelated questions like \textit{``Did you watch Tiktok last night?''} Such filtering is important to keep robots' pedagogical goal.

To prevent misuse of the robots, there should a preset time or schedule that students are aware of. For example, P4 and P6 suggested allowing teachers to \textit{``set a time frame''} or \textit{``routine.''}
P15 expressed their concerns if students have \textit{``unmoderated access, or unlimited access.''} P13 suggested a double authentication for power control. P4 proposed role-based control modes: \textit{``a teacher mode and a student mode''} – when in teacher mode, robots will only respond to the teacher and follow teacher's command; when in student's mode, robots will then respond to students.

In short, clear boundaries, expectation setting, and limited access are important when integrating robots into intercultural activities to make sure that robots are purposeful tools for intercultural learning rather than distractions.

\subsubsection{Integration into Teachers' Workflows}



To achieve the goal of supporting intercultural activities without overburdening teachers, robots should collaborate with teachers and fit into teachers' workflows, potentially by integrating into curriculum, offering teachers access to pre-developed activities, giving teachers enough control over robot, and providing after-interaction feedback. 

First, teachers suggested that activities should be integrated into curricula \textit{``with some standards attached to''} them (P14), so that they would be applicable to their own course materials (P10), and allow teachers to input their own materials for robots to deliver step-by-step (P4). If robot-mediated intercultural activities fall outside of curricula, they are likely to be abandoned easily. For example, P10 mentioned that AP classes have very tight curriculum, so they will prioritize them over soft skill development. Second, access to pre-developed activity resources was considered valuable. Teachers expected to have repositories or cloud-based libraries of intercultural activities, for different grade and student ability levels, which could be selected, updated, and customized as needed (P3, P4, P14) 
While some initial effort in learning and managing robots are inevitable, teachers emphasized that it should avoid \textit{``adding more work than the reward of it,''} especially the waste of time in tasks like programming, managing, and fixing it.

In addition, all teachers emphasized the importance of control over robot behavior. 
Robots were expected to act predictably, without \textit{``automatically turning on''} or \textit{``approaching students independently and starting talking to them''} (P10). During the activities, teachers would like to have real-monitoring and adjustments on interactions (P9, P14, P15).
Teachers also valued the post-activity feedback. P10 and P12 suggested robots to provide transcripts of language practice sessions, as a reference for both teachers and students; P7 suggested that robot could provide performance evaluation report. P9 highlighted the importance of using it to maintain the consistency when working with students, as P9 said: \textit{``I can see what was said and done, and maybe what I need to do for follow-up... I could see a negative part if the robot promised something like `a new notebook would be great for you', but I didn't know that. The child didn't get a new notebook, and was upset on the bus ride home.''} 

\subsection{Infrastructure Support and Requirements for Robot Use in School}
In this section, we present factors in a broader school-family support system that contribute to the long-term sustainable use of social robots for intercultural activities at school. 

\subsubsection{Curriculum Approval and Support}

For the sustainable use of robots in intercultural activities, school- and district-level support and guidance is critical, but often lacking. Teachers reported that intercultural competence development and social-emotional learning are not systematically embedded in the curriculum, but are \textit{ ``usually personal goals... There's not a lot of that built into the curriculum.''} (P14). With increasing cultural diversity in classrooms in recent years, P16 felt challenging and under-supported when dealing with it. As P16 said, \textit{``when they're all together, it gets rowdy... When I try to get help at school, it seems like they blame me.''} 
Meanwhile, the curriculum demands sometimes limit opportunities to focus on intercultural and social emotional development, because their \textit{``school district is very focused on curriculum, so [teachers] don't have a lot of time to focus on the other skills''} (P8). 

When considering using social robots to support intercultural activities, teachers pointed out that approval from the school board and district would be mandatory to include it into curriculum (P16), and a rigorous review would surely be required \textit{``even though it is perfectly safe''} (P17). Some teachers doubt the district would support the use of robots, given the broader resistance to AI and related technologies in classroom (P8). P17 mentioned that robots for intercultural activities might not meet curriculum standards, since they do not directly contribute to the learning of technical knowledge as other tools in computer science classes do. 
Beyond worries on curriculum approval, financial sustainability is always a concern – that is another reason why continuous support and approval from district is so necessary. 
Shot-term funding was not enough: P9 shared their experience of receiving a grant for initial use of some robots and tablets, but facing the problem of maintenance or replacement costs after funds ran out. Only if the robot's price is affordable and there is sufficient funding and support from school- and district- levels, the practice could be sustainable (P1, P7).


    





\subsubsection{Ethical Considerations and Compliance}
When social robots interact with minors, ethical concerns naturally occur. Within the intercultural context, they become even more complex. 
Participants expressed uncertainty about what information robots might record and how it would be stored, accessed, or disclosed. Teachers worried that robots' \textit{recording} functions will add risks and be controversial when interacting with minors (P15). 
If recording is a prerequisite for use, P17 pointed out that, \textit{``it should be very clear if a device is on and recording''}, and \textit{``be very obvious to tell when something is recorded or not.''} Teachers also worried about the \textit{hacking risks} of collected data where sensitive student information could be stolen and misused by \textit{``identity thieves''} (P16). Teachers further emphasized the importance of compliance with institutional and legal standards, such as the Family Educational Rights and Privacy Act (FERPA). As P17 explained, we must make sure that \textit{``we're not having any kind of legal issues, any kind of FERPA violation.''} and \textit{``we have assurance that, security-wise, everything is ready for use.''}

More importantly, there are risks associated with cultures. P17 reported some examples of teachers who lost job after being accused of using inappropriate culture-sensitive words. As reported by P17, \textit{``in some regions, social emotional learning is seen as political. People are concerned these kinds of things being inappropriate for students, even if it's not inappropriate.''} Similarly when considering robot-involved intercultural activities, it is critical to make sure robots respect cultural diversity while also meeting compliance with the political and cultural norms of the local context.


\subsubsection{Parents' Engagement and Transparency}
Parents are key stakeholders in K-12 students' development. Given that culture-related topics can sometimes be sensitive, transparency, consent, and open communication between schools and families are essential.
Teachers anticipated that parents would likely to have concerns about safety and security, and doubts about cultural topics (P5, P17,P16, P15). 
Some participants described the importance of introducing robots to parents early on to gain their consent. P5 explained the need to \textit{``do a formal introduction to students, get their agreement, brainstorm on the rules, and inform parents about everything, including the education goals and the reason for taking a robot into the classroom''}. P17 suggested developing a standard script to help schools communicate with families. In their view, such a script, should explain the robot's role, outline its functions, and address questions about security and privacy.





Teachers also noted that parents' concerns could arise during ongoing activities. Here, ensuring the transparency of the activities could help. As P15 proposed, \textit{making a script of the activity available to parents} could enhance transparency about the classroom interactions. While P15 anticipated fewer difficulties in relatively open-minded area, it could still be a precautionary measure.



\section{Discussion}
\begin{figure*}
\includegraphics[width=0.8\textwidth]{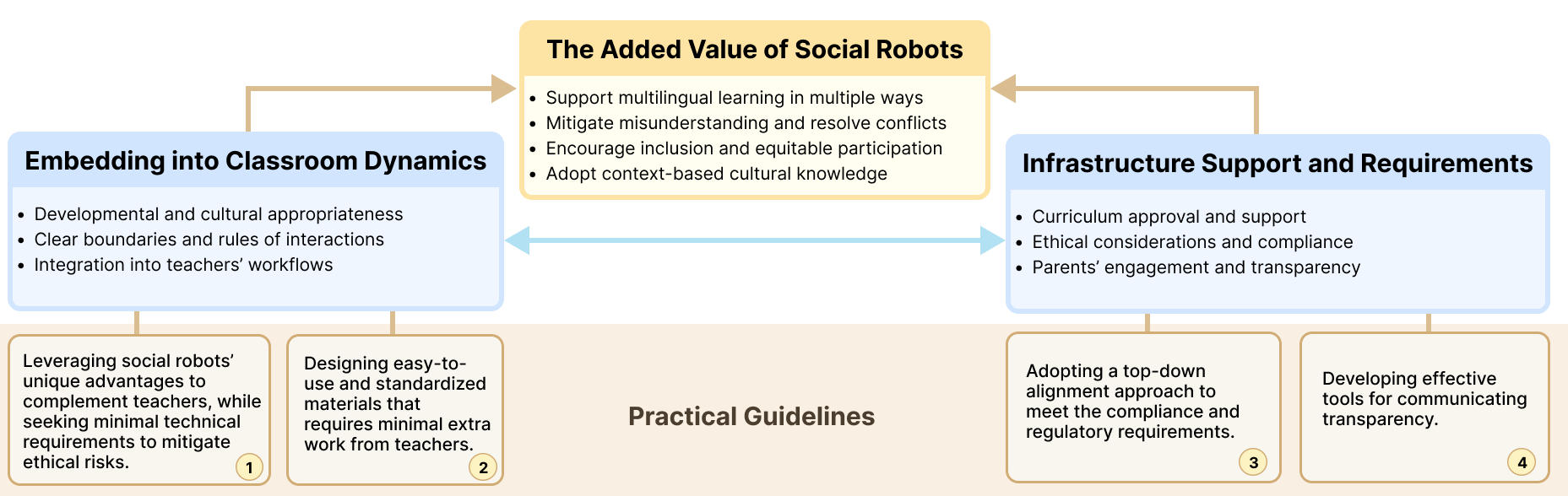}
    \vspace{-1.5em}
  \caption{\textbf{Incorporating Social Robots into School to Support Intercultural Competence}}
  \Description{This diagram shows how the qualitative findings inform a set of practical guidelines for developing HRI systems to support students' intercultural competence. ``Embedding into Classroom Dynamics'' and ``Infrastructure Support and Requirements'' are identified as necessary conditions for achieving ``The Added Value,'' while the guidelines provide actionable steps for implementation in school contexts.}
  \label{fig:guidelines}
  \vspace{-2em}
\end{figure*}
These findings provide insights into teachers' perspectives on using social robots to support students' intercultural competence development at school. We now discuss how these findings inform a set of actionable guidelines in designing social robots for intercultural competence development, with the ultimate goal of \textbf{integrating} them into schools. As shown in Fig. \ref{fig:guidelines}, ``Embedding into Classroom Dynamics'' and ``Infrastructure Support and Requirements'' are conditions for achieving ``The Added Value'' of social robots, and the guidelines below provide actionable steps for achieving them.




\textbf{Social robots have unique values in supporting intercultural activities}, which could be categorized into three aspects: knowledge, functionality, and social aspects.
In terms of knowledge, 
previous research has highlighted the importance of robots' bilingual abilities in engaging linguistically diverse children equally \cite{kim2021designing}. In addition to language, our study finds that context-based cultural knowledge is valuable in school-based activities, complementing teachers in providing appropriate intercultural explanations, both as a learning resource and as a mediator in conflict scenarios. 
In terms of the functionality, robots can play diverse flexible roles \cite{ekstrom2022dual}  
and offer additional help in managing the class when there is insufficient number of teachers. 
This kind of diversity and flexibility in roles is particularly valuable in intercultural interactions, adapting to culturally diverse students' needs (e.g., multilingual learners \cite{louie2022designing}), where teachers might feel overwhelmed.
In terms of the social aspect, robots demonstrate some advantages thanks to their unique social identity as an unfamiliar third-party with a non-judgmental stance. 
Our participants varied in their cultural and linguistic background and all had experience working with students from diverse and non-dominant communities. Still, they  described uncertainty and limitation when interpreting cultural differences outside their knowledge. In this situation, social robots are perceived as a neutral other, helping create space for cultural exchange that teachers find difficult to initiate alone. Prior studies find students are more likely to open up to a social robot compared to a human on personal topics \cite{abbasi2022can}. In intercultural contexts, a robot's presence could be particularly useful to create a psychologically safe space and foster meaningful participation in intercultural activities. 
Nevertheless, this does not imply that social robots are inherently neutral.
For example, differences in appearance and behavior can lead to the racialization of robots and influence people's interactions with them. \cite{strait2018robots, sparrow2019robots, bartneck2018robots}.
This further reminds us to be mindful of unintentionally embedded biases in robot design and the positionality of different stakeholders, and to be cautious about simply transferring human teachers' strategies to social robots.

Meanwhile, 
\textbf{\textit{ethical and safety concerns}} need extra consideration. While previous research discussed the physical and psychological safety concerns around the children-robot interaction in educational environments \cite{you2025investigating}, we find that 
safety and privacy
is a concern shaped by complex environmental and system-level factors, which greatly influence stakeholders' willingness of adoption \cite{alfredo2024human}. 
This tension is amplified by the fact that researchers are able to use many sensors and collect rich high-quality data during experiment, which is hard to sustain in real classroom deployment due to cost and infrastructure \cite{woo2021use}.
It 
highlights \textbf{two distinct orientations in HRI: mechanism-oriented and practice-oriented research}. Mechanism-oriented research is usually conducted under highly controlled conditions, allowing for the substantial data collection to explore the fundamentals of human-robot interaction and then on that basis improve robots' behaviors and performance. However, practice-oriented research is defined by identifying necessary compromises under restrictions \cite{woo2021use}: data collection is limited, privacy concerns are significant, schools and parents set strict boundaries, and money is often tight. In this context, the goal shifts from theoretical or scientific breakthroughs to developing solutions that are feasible and ethical under real-world constraints.
This leads to the \textbf{first guideline} – \textbf{\textit{Leveraging social robots' unique advantages to complement teachers, while seeking minimal technical requirements to mitigate ethical risks}}.








Introducing new technologies inevitably brings extra work at the beginning, which is a general concern when teachers already have heavy workload \cite{ahumada2025robots, reich2016robots}. 
While some initial effort is inevitable, robots should bring long-term benefits and make their work easier other than harder. Teachers expect robot-mediated intercultural activities to be closely aligned with their daily teaching objectives and workflows. Thus, they should be embedded and able to integrate into classroom routines, instead of being a single separate activity. In addition, if robots and the supporting materials are plug-and-play and used immediately, it will save a considerable amount of time and effort for teachers. For example, block programming tools could lower teachers' burden in using robotic kits \cite{ali2023exploring}. Meanwhile, teachers value a certain level of agency and autonomy in selecting the most appropriate materials for their students, such as difficulty level, cultural themes, or modes of interaction. This leads to our \textbf{second guideline} – \textit{\textbf{Designing easy-to-use and standardized materials that requires minimal extra work from teachers}}.

To introduce social robots for intercultural activities in school, decision-making authority is held at the school and district levels. Given the sensitivity of intercultural topics and the safety concern of social robots, teachers cannot simply introduce robots in classrooms for long-term interactions without official approval. To make the work sustainable and meaningful, the integration of social robots has to be in the curriculum. This way, the activities are connected to learning goals and supported over time. Such needs point to the adoption of a top-down approach, where districts and schools can establish clear instructions, regulatory compliance measures, and operational procedures. Such kind of institutional endorsement will provide legitimacy, address safety and privacy concerns, and ensure the use of social robots in intercultural activities qualifies the requirements of broader education policy \cite{johnston2023privacy}. This leads to the \textbf{third guideline} – \textbf{\textit{Adopting a top-down alignment approach to meet the compliance and regulatory requirements.}}





In addition, transparency is a key factor to build acceptance and trust. HRI researchers have some exploratory work on designing robot behaviors to make users aware of the decision processes behind robot actions \cite{cumbal2024speaking}. For intercultural activities in educational contexts, transparency should also address more than the people who directly interact with the robots. Most stakeholders, including students, teachers, parents, school administrators, are not technology experts, and may also have concerns when it involves culture-sensitive content. They need a clear understanding of the activities and robots' role and limitations in supporting such activities. Teachers must know the teaching goal, robot functions and control, students need clarity about the rules and boundaries of interaction, parents worry about data collection and privacy protection, and school administrators care about compliance and alignment with school values. There is a need for communication on transparency and consensus between all different stakeholders. This leads to the \textbf{fourth guideline} – \textit{\textbf{Developing effective tools for communicating transparency}}.

\subsection{Practical Guidelines} 
\label{guidelines}

\textit{\textbf{(1) Leveraging social robots' unique advantages to complement teachers, while seeking minimal technical requirements to mitigate ethical risks.}}
We suggest making better use of robots' advantages in intercultural activities. Specifically, the design should explicitly target tasks where robots' strengths in knowledge, functionality, and social presence that complement teachers' expertise without taking over teachers' core roles. 
Meanwhile, taking a practice-oriented approach, we recommend prioritizing the exploration of the minimal technical requirements for effective operation in classrooms to reduce ethical risks and financial burdens by avoiding unnecessary data collection, lowering costs, and ensuring that robots remain manageable and accessible for teachers and schools.




\noindent \textit{\textbf{(2) Designing easy-to-use and standardized materials that requires minimal extra work from teachers.}} 
We suggest designing easy-to-use and standardized materials with options for teachers to select from. It could be an online bank or cloud where standardized templates, guides, and resources are provided, while teachers should still able to make choices such as selecting the right difficulty, cultural themes, or interaction style for their students. It helps teachers feel in control, without overburdening them.

\noindent \textit{\textbf{(3) Adopting a top-down alignment approach to meet the compliance and regulatory requirements.}}
We suggest that HRI researchers adopt a top-down alignment approach to ensure that the use of robots in intercultural activities meet existing compliance and regulatory frameworks, and also build trust among stakeholders. In practice, this means collaborating with school administrators, district policymakers, and regulatory bodies to define appropriate guidelines and curriculum. It also involves developing clear operational protocols – such as how robots should be introduced into classrooms, how teachers and students should interact with them, and how usage should be monitored and evaluated. By taking a top-down approach, researchers can help create a supportive environment for the responsible and sustainable adoption of robots for intercultural activities in schools.



\noindent \textit{\textbf{(4) Developing effective tools for communicating transparency.}}
We encourage HRI researchers to develop effective tools for communicating and demonstrating transparency. Researchers should collaborate with experts from multiple domains in the design of such tools, including, but not limited to, technical, educational, privacy and security, and ethics and legal expertise. In addition, it should have different emphasis when targeting different stakeholders. In general, the goals are to help teachers use the robots for intercultural activities effectively, to clearly communicate boundaries and rules to students, to address parents' concerns about privacy and safety, and to explain the its value to schools and administrators.

\section{Limitations and Future Work}
Our study includes only K-12 teachers based in USA, thus findings may not reflect other cultural contexts; Using Haru as the probe may have unintentionally narrowed ideation toward its current morphology and capabilities; There is a lack of direct engagement with students and their families from different cultural backgrounds, which we identify as an important direction for future work. 

\section{Conclusion}
This study aims to understand the opportunities and practical considerations of integrating social robots in schools to support students' development of intercultural competence. Through co-design workshops with teachers, we first identified the added value of robots in supporting intercultural competence at school, including multilingual scaffolding, culturally situated knowledge and a perceived neutral third-party in conflict resolution; then classroom embedding requirements, including developmental and cultural appropriateness, clear role boundaries, and workflow integration; and finally infrastructure conditions for sustainable use, including curriculum support, ethics and compliance, and transparent communication. Based on the findings, we proposed a set of practical guidelines for HRI community to design social robots that are are adoptable, maintainable, and compliant within school contexts.

\begin{acks}
We thank Honda Research Institute JP for funding this research. We thank the participants and reviewers for their valuable contributions. We appreciate the support of other RHouse Lab members.
\end{acks}

\bibliographystyle{ACM-Reference-Format}
\bibliography{0-References}

\appendix

\end{document}